\documentclass[aps, pra, onecolumn,amsmath,amssymb,10pt,notitlepage]{revtex4-2}

\usepackage{graphicx}
\usepackage{dcolumn}
\usepackage{bm}
\usepackage{braket}
\usepackage{amsmath}
\usepackage[colorlinks=true,bookmarks=false,allcolors=blue]{hyperref}

\usepackage[final]{changes}
\newcommand{\revadd}[1]{\added{#1}}
\newcommand{\revdel}[1]{\deleted{#1}}
\newcommand{\revrep}[2]{\replaced{#1}{#2}}

\begin{document}
\title[Random Metasurfaces for Photonic Qudit Tomography]{Can Randomly Structured Metasurfaces Be Used for Quantum Tomography of High-Dimensional Spatial Qudits?}
\author{Yuming Niu}
\author{Kai Wang}
\email{k.wang@mcgill.ca}
\affiliation{Department of Physics, McGill University, 3600 Rue University, Montreal, Quebec, H3A 2T8, Canada}
\date{\today}

\begin{abstract}
Reconstructing the density matrix of the quantum state of photons through a tomographically complete set of measurements, known as quantum state tomography, is an essential task in nearly all applications of quantum science and technology, from quantum sensing to quantum communications. Recent advances in optical metasurfaces enable the design of ultrathin nanostructured optical elements performing such state tomography tasks, promising greater simplicity, miniaturization, and scalability. However, reported metasurfaces on this goal were limited to a small Hilbert dimension, e.g., polarization qubits or spatial qudits with only a few states. When scaling up to higher-dimensional qudit tomography problems, especially those involving spatial qudits, a natural question arises: whether a metasurface with randomized nanostructures is sufficient to perform such qudit tomography, achieving optimal conditions. In this work, we attempt to answer this question through a set of numerical experiments with random metasurfaces, utilizing large-scale simulations of over \revrep{16,000}{ 11,000} distinct metasurfaces each exceeding 200 wavelengths in size. We show that, with sufficient redundancy in the number of detectors, random metasurfaces perform reasonably well in quantum photonic spatial qudit tomography encoded in Hermite-Gaussian states for up to approximately 10 states. Furthermore, we discuss additional considerations for optimizing metasurfaces in multiphoton cases. Our work opens a pathway toward computationally efficient, miniaturized, and error-tolerant quantum measurement platforms.
\end{abstract}


\maketitle 

\section{\label{sec:Intro} Introduction}

One key advantage of using photons in quantum information is that they have multiple internal degrees of freedom, including polarization, spatial, and spectral state. These intrinsic degrees of freedom allow for the encoding of information beyond a two-dimensional Hilbert space (qubit), as a single-photon state can be treated as a \emph{qudit} described in a higher-dimensional Hilbert space~\cite{erha20NatRevPhys}. Decoding the embedded information on photons in terms of their density matrices is through quantum state tomography (QST)~\cite{jame01Phys.Rev.A, thew02Phys.Rev.Aa}, where one performs a tomographically complete set of quantum measurements and hence reconstructs the unknown state using the measurement results. The implementation of quantum state tomography conventionally relied on bulky classical optics to create such a set of measurements~\cite{jame01Phys.Rev.A, thew02Phys.Rev.Aa, agne11Phys.Rev.A, bent15Phys.Rev.X, kues17Nature, sosa17Phys.Rev.Lett., curi19Photon.Res.PRJ, zia23Nat.Photon.}. 
However, as the number of free parameters in the density matrix scales quadratically with the Hilbert dimension and even faster with the number of photons, conventional quantum state tomography approaches suffer from poor scalability and bulky and error-prone setups. Therefore, a more scalable paradigm for multiphoton qudit measurement is urgently required for advanced quantum photonic experiments.

Recent progress in the field of flat optics promises to replace and even outperform conventional optical elements with nanostructured metasurfaces~\cite{ yu14NatureMater, gene17OpticaOPTICA, kuzn24ACSPhotonics, roqu25ACSPhotonics}. Optical metasurface, an ultrathin element composed of carefully designed subwavelength structures, can manipulate light in all degrees of freedom with ultimate miniaturization. 
While most of the metasurfaces were developed for applications in classical optics such as meta-lenses~\cite{khor16Science} and meta waveplates~\cite{arba15NatureNanotech}, there have been emerging efforts to use metasurfaces for the generation~\cite{stav18Science, li20Science, sant22Science, nemi25}, manipulation~\cite{wang18Science, geor19LightSciAppl, lung20ACSPhotonics, yous25Sciencea}, and measurement~\cite{wang18Science,wang23NanoLett.a,lung24ACSPhotonics,an24NatCommun,ren24Phys.Rev.Res.} of quantum state of light~\cite{soln21Nat.Photonics, wang22PhysicsToday, ding23MaterialsToday,zhan24APLQuantum}. 
In particular, many works employed metasurfaces in conjunction with single-photon counting sensors to measure single- and multiphoton quantum states~\cite{wang18Science,lung24ACSPhotonics,an24NatCommun,wang23NanoLett.a}. In these works, the measurements or generalized measurements required for quantum state tomography are realized in parallel, all generated by a monolithic metasurface, promising better scalability, robustness against errors, and miniaturization.
However, to date, these efforts have been limited to a small Hilbert dimension, as they use either the polarization degree of freedom (qubits)~\cite{wang18Science,lung24ACSPhotonics, an24NatCommun} or utilize only a tiny number of spatial states~\cite{wang23NanoLett.a}. It remains an open question whether optical metasurfaces can facilitate a high-dimensional quantum state tomography of photons.

This work focuses on multidimensional quantum photonic states encoded in the spatial degree of freedom. It is well-known that spatial quantum state tomography problems are closely related to classical phase, amplitude, and coherence retrieval. Indeed, in classical optics and photonics, numerous methods have been developed for retrieving classical light field information. In particular, many of them are based on using sufficiently random structures, such as for generating speckles to derive the underlying phase and coherence information~\cite{anan07Opt.Lett.OL, anti18OpticaOPTICA, hori14Opt.Lett.}. On the other hand, in quantum information, it is also known that random projections can be used for quantum state tomography~\cite{mirh14Phys.Rev.Lett., howl16Phys.Rev.X, pg21Phys.Rev.A}. Therefore, it is a natural question whether we can use a randomly shaped optical metasurface for quantum state tomography of the spatial qudit states of photons. This question is highly nontrivial, as the material and photonic structure configuration of the metasurface impose limitations on the achievable measurements, where a randomized metasurface structure is not equivalent to randomized measurements. The random nature of the structure and the possible cross-talks between light fields in different meta-atoms make it challenging to investigate this problem through the typically employed small unit cells under periodic boundary conditions.

In this work, we aim to investigate if metasurfaces with random freeform structures are capable of performing tomography.
Without loss of generality, we choose Hermite-Gaussian (HG) spatial states as the basis. Instead of using the conventional metasurface design approach, where only a few structure parameters of a preselected geometry are optimized within a small unit cell, we generate a large number of large-scale, freeform random structures under different settings and characterize their performance with numerical simulations of the entire metasurface. 

We emphasize that the exploration of whether a randomly shaped metasurface works well for quantum state tomography is beyond a curiosity question; there are significant implications in practical applications. Using random structures can substantially save the time and computational resources required for traditional inverse design with a huge parameter space. More importantly, random structures are naturally robust against fabrication errors. Small mutations of the metasurface structure from fabrication would indeed change the transformation on the spatial photon, but as long as the transformation does not become poor conditioned for reconstruction, one can always run a calibration to use the actual transformation in the experiment. If a randomized structure works in certain regime, it already guarantees that small structural deviations will not significantly change the condition of the inverse problem. 
Furthermore, studying this problem through numerous random tests will provide insights into the global features of this problem, which is also beneficial for the search for potential inverse design strategies, as the latter are typically based on local optimizations.

\section{\label{sec:HGQST}
Metasurface-Based Hermite-Gaussian Spatial Qudit Tomography}

\begin{figure}[t]
\includegraphics[width=0.8\textwidth]{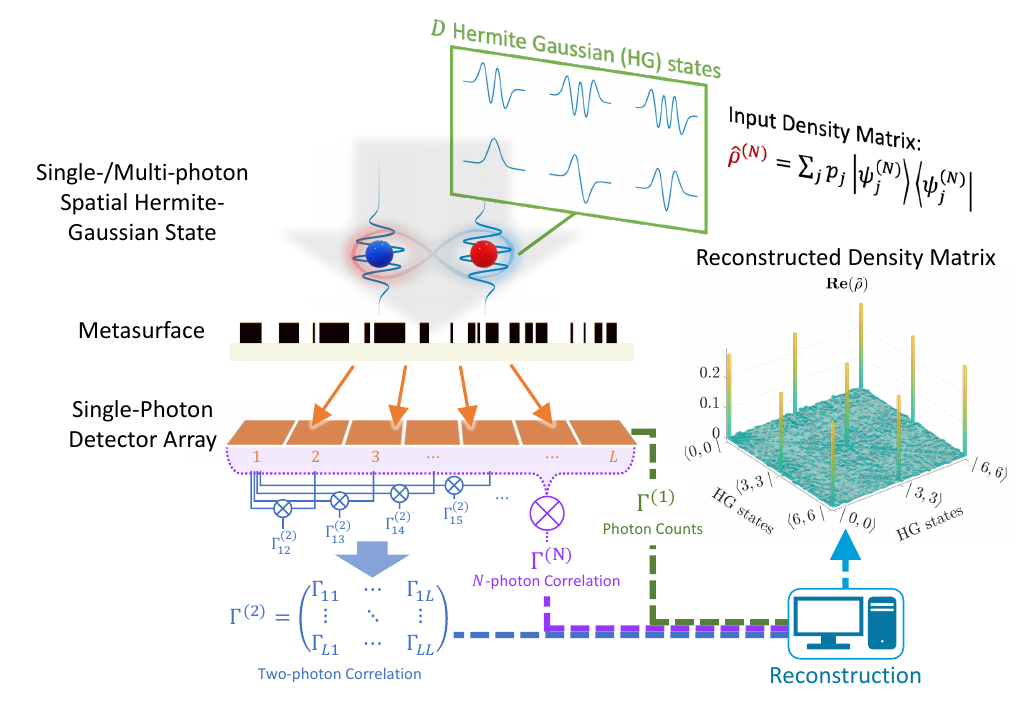}
\caption{\label{fig:schematic}Schematic of the spatial qudit tomography of photons with metasurfaces. Single- or multi-photon spatial state encoded under a finite $D$-dimensional Hermite-Gaussian (HG) basis is refracted by a metasurface onto a single-photon detector array, which measures average photon counts $\bm{\Gamma}^{(1)}$ of each pixel for a single-photon state reconstruction or an $N$-fold correlation  $\bm{\Gamma}^{(N)}$ among pixels for an $N$-photon state. The reconstruction algorithm recovers the density matrix of the input knowing the transformation of the device.}
\end{figure}

We start by illustrating the schematic of the single- or multiphoton qudit tomography using a piece of metasurface in Fig.~\ref{fig:schematic}. The photons are prepared in an unknown spatial state with a fixed photon number $N$ characterized by a density matrix $\hat{\rho}^{(N)}$. The metasurface diffracts the transmitted photons in a way such that their probability distribution and the correlations on the single-photon detector array placed in the far field contain sufficient information to reliably derive the input unknown state's density matrix $\hat{\rho}^{(N)}$.

Without loss of generality, we set Hermite-Gaussian (HG) spatial states as the basis of interest for our multidimensional tomography problem. At the waist, the spatially dependent probability amplitude of the one-dimensional (1D) HG state of order $m$ can be expressed as
\begin{equation}
    \ket{m}\equiv\ket{\mathrm{HG}_m (x)}= {a_0}
\cdot {\mathbb{H}_m}\left( {\sqrt 2 \frac{x}{{w_0}}} \right) \cdot  \exp \left( { - \frac{{{x^2}}}{{w_0^2}}} \right),
 \label{HG}
\end{equation}
where $\mathbb{H}_m(x)$ is the Hermite polynomial of order $m$,  $w_0$ is the waist radius of the fundamental order ($\ket{0}$~state), and $a_0$ is an amplitude normalization constant. HG spatial states form a semi-infinite, orthogonal, and complete basis for monochromatic beams of the waist $w_0$. States with higher HG orders have finer spatial details, and are analogous to states associated with higher Fourier frequencies. As a common assumption, a measurement system handles details up to a certain extent. Thus, it is natural to utilize a finite Hilbert space of HG states by truncating them to some finite order $m=0,1,\dots, D-1$. Hence, the Hilbert dimension for a single-photon state in this representation is $D$. 

For a single-photon pure state in this basis, the state vector is written as 
\begin{equation}
    \ket{\psi^{(1)} }=\sum_{m=0}^{D-1} c_m \ket{m}.
\end{equation}
Note that the single-photon state vector $\ket{\psi^{(1)}}$ can only represent a pure state; a single-photon state's density matrix $\hat{\rho}^{(1)}$ represents more general cases including mixed states, which can be understood through the spectral decomposition
\begin{equation}
    \hat{\rho}^{(1)}=\sum_{j} p_j \ket{\psi^{(1)}_j}\bra{\psi^{(1)}_j},
\end{equation}
where $\hat{\rho}^{(1)}$ is a $D\times D$ Hermitian matrix with non-negative eigenvalues, which can be written as an incoherent combination of pure states $\ket{\psi^{(1)}_j}$, each with a probability of $p_j$. 

Likewise, in the case of a fixed photon number $N>1$ (i.e., multiphoton), if there is some other degree of freedom that can distinguish (label) the photons, we can denote an $N$-photon pure state in such a truncated HG basis as
\begin{equation}
    \ket{\psi^{(N)} }=\sum_{m_1, \dots, m_N=0}^{D-1}c_{m_1, \dots, m_N}\ket{m_1, \dots, m_N},
\end{equation}
which is a vector with $D^N$ elements.
Accordingly, the $N$-photon density matrix is
\begin{equation}
    \hat{\rho}^{(N)}=\sum_{j} p_j \ket{\psi^{(N)}_j}\bra{\psi^{(N)}_j},
\end{equation}
which is a $D^N \times D^N$ Hermitian matrix with non-negative eigenvalues. 

Our spatial qudit tomography aims to reconstruct the fully-unknown density matrix, $\hat{\rho}^{(N)}$ with a fixed photon number $N$, which includes all information on amplitude, phase, coherence, and quantum entanglement.
In general, one would need to obtain at least $D^{2N}$ measurements as there are $D^{2N}$ independent real-valued unknowns in $\hat{\rho}^{(N)}$ without considering normalization.

The metasurface's functionality is to create a well-conditioned set of sufficiently many measurements that can well reconstruct $\hat{\rho}^{(N)}$. Importantly, these measurements must be realistic for the detectors or image sensors, which typically perform intensity or photon-counting measurements. In this work, we consider a sensor that is a 1D array of single-photon detectors with $L$ pixels, where each pixel is indexed by $l\in\{1,\dots,L\}$. Hence, such measurements correspond to average photon counts in the single-photon case and $N$-fold correlations across the pixels in the $N$-photon case with $N>1$. Such correlation measurements can be described by
\begin{equation}
    \Gamma^{(N)}_{l_1, \dots, l_N}=\mathrm{Tr}\left(\hat{F}^{(N)}_{l_1, \dots, l_N} \hat{\rho}^{(N)}\right),
\end{equation}
where $\Gamma^{(N)}_{l_1, \dots, l_N}$ is the $N$-fold correlation that corresponds to the probability of $N$ photons indexed by $1,\dots,N$ simultaneously clicking pixels $l_1, \dots, l_N$, respectively. Here, $\hat{F}^{(N)}_{l_1, \dots, l_N}$ is the positive operator-valued measure (POVM)\revrep{. In quantum measurement theory, the outcome of any physical measurement can be represented by a set of POVMs, which generalize the concept of projective measurements. In the context of our metasurface-based tomography, each detector pixel corresponds to one such POVM element, determined by the metasurface’s optical transformation. }{that describes this generalized measurement, which is determined by the metasurface's diffraction properties. }We note that $\Gamma^{(N)}$ is basically a correlation matrix of dimension $L^N$. Generally, there are primarily two types of single-photon detectors, one that cannot resolve the number of photons in each detection event, known as click detectors, and the other is photon-number-resolving detectors~\cite{hadf09NaturePhoton}. Click detectors, most commonly used, can only measure fully nonlocal correlations. Here, for the simplicity of argument, we assume the detectors are photon-number-resolving detectors. Nevertheless, our scheme works well for click detectors; our two-photon reconstruction example in Sec.~\ref{sec:Rec} will be based on click detectors, and we will also provide detailed discussions on the additional consideration of click detectors in Sec.~\ref{sec:diss}. 
For an array of $L$ detectors, there are in total $L^N$ measurements for $N$ distinguishable photons. In the single-photon case of $N=1$, the measurement reduces to single-photon counting at each pixel, i.e., 
$\Gamma^{(1)}_{l}=\mathrm{Tr}\left(\hat{F}^{(1)}_{l} \hat{\rho}^{(1)}\right).$

Importantly, having more measurements than the number of unknowns, i.e., $L^N\geq D^{2N}$, is a necessary yet not sufficient condition. After this well-posedness condition is met, the key to running a reliable reconstruction is that the set of POVMs must be well chosen in a way that the reconstruction is a well-conditioned inverse problem. The metasurface design mainly concerns this factor. It is critical to note that such a requirement does not target specific POVMs; rather, it is based on an overall measure of all the POVMs. Thus, the choice is far from unique. Such a requirement is the key underlying principle supporting the conjecture that a random metasurface might work. Indeed, there should be many well-conditioned choices of the set of POVMs; in practice, one can run a calibration of the metasurface-detector system to characterize each POVM before performing tomography measurements. After the calibration is considered, the only factor that determines the performance of the reconstruction is how well-conditioned it is to inversely derive the input quantum state based on the outcomes associated with the POVMs.
To mathematically formulate such an inverse-condition consideration, we can denote the entire set of measurements using one linear equation
\begin{equation}
    \bm{\Gamma}^{(N)} = \mathbf{M}^{(N)} \bm{\rho}^{(N)},\label{eq_instrument_matrix_N}
\end{equation}
where $\bm{\Gamma}^{(N)}$ is a column vector flattened from the correlation matrix ${\Gamma}^{(N)}$ mentioned above, and likewise $\bm{\rho}^{(N)}$ is the column-vector form of $\hat{\rho}^{(N)}$. Here, $\mathbf{M}^{(N)}$ is the so-called instrument matrix connecting the elements in the measured density matrix to all the correlation measurements; each row of $\mathbf{M}^{(N)}$ corresponds to each POVM and can be obtained by reshaping $\hat{F}^{(N)}_{l_1, \dots, l_N}$ into a row vector. The qudit tomography problem can be phrased as finding the flattened input density matrix, $\bm{\rho}^{(N)}$, given the noisy measurement vector, $\tilde{\bm{\Gamma}}^{(N)}$, which is a linear inverse problem characterized by Eq.~\eqref{eq_instrument_matrix_N}. The reconstruction is essentially computing the (pseudo) inversion of matrix $\mathbf{M}^{(N)}$, i.e., the reconstructed state being $\bm{\tilde{\rho}}^{(N)}={\mathbf{M}^{(N)}}^{-1}\bm{\Gamma}^{(N)}$, although the inversion may not be directly computed.
Therefore, the design requirement of the metasurface is centered around how well the inverse condition is for $\mathbf{M}^{(N)}$.

Based on Eq.~\eqref{eq_instrument_matrix_N}, we can now concretely describe the inverse condition in such a reconstruction problem using $\mathbf{M}^{(N)}$. After the basic condition of $L^N\geq D^{2N}$ is met, the consideration lies in the noise amplification when one inverts the process. It is known that a worst-case-scenario noise amplification of such linear processes can be measured by the so-called condition number~\cite{golu13} defined as $\kappa(\mathbf{M}^{(N)})\equiv {\|\mathbf{M}^{(N)}\|}{\|{\mathbf{M}^{(N)}}^{-1}\|}$. In the case of matrix $2$-norm, we have
\begin{equation}
    \kappa(\mathbf{M}^{(N)})= {\|\mathbf{M}^{(N)}\|}_2{\|{\mathbf{M}^{(N)}}^{-1}\|}_2 =\frac{\sigma_{\max}(\mathbf{M}^{(N)})}{\sigma_{\min}(\mathbf{M}^{(N)})}. \label{cond_def}
\end{equation}
where $||\cdot||_2$ denotes the matrix $2$-norm, and $\sigma_{\max}$, $\sigma_{\min}$ refers to the maximum and minimum singular values, respectively. In our study, we primarily use the condition number with respect to the matrix 2-norm to evaluate the performance of the simulated metasurfaces. 

To see how the condition number is related to the worst-case noise amplification in the tomography, suppose a measurement error, $\bm{\epsilon}$ perturbs the true signal $\bm{\Gamma}^{(N)}$, yielding $\tilde{\bm{\Gamma}}^{(N)} = \bm{\Gamma}^{(N)} + \bm{\epsilon}$ in the measurement process. Hence, the inversion would give an error of ${\mathbf{M}^{(N)}}^{-1}\bm{\epsilon}$ from the true input, ${\mathbf{M}^{(N)}}^{-1}\bm{\Gamma}^{(N)}$. By definition of the matrix $2$-norm and Eq.~\eqref{cond_def}, there is
\begin{equation}
    \kappa(\mathbf{M}^{(N)})= \max _{\bm{\Gamma}^{(N)}, \bm{\epsilon} \ne \bm{0}}\left( \frac{R_\text{rec}}{R_\text{mea}} \right),\label{cond_rel_err}
\end{equation}
where we define $R_\text{rec} \equiv \|{\mathbf{M}^{(N)}}^{-1}\bm{\epsilon}\|_2 / \|{\mathbf{M}^{(N)}}^{-1}\bm{\Gamma}^{(N)} \|_2$ and $R_\text{mea} \equiv \|\bm{\epsilon}\|_2/\|\bm{\Gamma}^{(N)}\|_2$ as the relative error of the reconstructed result and measurement, respectively. Eq.~\eqref{cond_rel_err} shows that the condition number $\kappa$ is the maximum of the relative reconstructed state error over the relative measurement signal error; in other words, the change in relative error from reconstruction is bounded by the condition number, i.e. 
$\kappa(\mathbf{M}^{(N)}) \ge R_\text{rec}/R_\text{mea} \ge 1/\kappa(\mathbf{M}^{(N)}),$
where the second inequality comes from swapping the signal and the error in Eq.~\eqref{cond_rel_err}, since signal and error have no essential difference in the analysis above.
As the reconstruction process is essentially based on a multiplication of ${\mathbf{M}^{(N)}}^{-1}$ onto the measured signal, a small condition number $\kappa$ would ensure that noises in the detectors' readout are not excessively amplified, and thus we call the instrument matrix $\mathbf{M}^{(N)}$ to be \textit{well-conditioned}.
In other words, this condition number characterizes how accurately different input states can be differentiated by this metasurface. 

We note that the relative error, e.g., at the sensor, $R_\text{mea}$, is closely related to the signal-to-noise ratio (SNR) 
\begin{equation}
    \mathrm{SNR} \equiv \frac{\mathrm{mean}( \bm{\Gamma}^{(N)})}{\mathrm{std}(\bm{\epsilon})}\label{SNR_meas},
\end{equation}
defined for the noise-free signal $ \bm{\Gamma}^{(N)}$, and a zero-mean noise vector $\bm{\epsilon}$ that has a standard deviation of $\mathrm{std}(\bm{\epsilon})$.
As a result, the condition number reflects the change of the signal-to-noise ratios during reconstruction. We can see that the best metasurface for this task is those with minimal $\kappa(\mathbf{M}^{(N)})$. 

While the case of an $N$-photon tomography problem comes with a huge-size instrument matrix $\mathbf{M}^{(N)}$ as its dimension is $L^N\times D^{2N}$, it is not always necessary to formulate the reconstruction condition through an $N$-photon case. This is because a linear metasurface is a classical object, and the POVMs that it generates for $N$-photon state is trivially connected to the POVMs for single-photon states. More specifically, assuming the metasurface performs the same linear transformation for each photon, there is
\begin{equation}
\hat{F}^{(N)}_{l_1, \dots, l_N} = \hat{F}^{(1)}_{l_1} \otimes \dots \otimes \hat{F}^{(1)}_{l_N},
\end{equation}
where $\otimes$ denotes the Kronecker product.
Likewise, for the instrument matrix, we have
\begin{equation}
\mathbf{M}^{(N)}= {\mathbf{M}^{(1)}}^{\otimes N}. \label{tensor_prod_M}
\end{equation}
Consequently, there is $\kappa (\mathbf{M}^{(N)})=\kappa^N (\mathbf{M}^{(1)})$. Therefore, if there is no additional consideration for the $N$-photon measurement, we can simply use the single-photon instrument matrix $\mathbf{M}^{(1)}$ and the corresponding single-photon-case condition number $\kappa(\mathbf{M}^{(1)})$ as a generic measure of this inverse problem. Therefore, in the numerical tests we run in this paper, we primarily focus on using $\kappa(\mathbf{M}^{(1)})$ as the measure. Nevertheless, we will also show examples of two-photon state reconstruction and discussions on special cases where the multiphoton instrument matrix has to be considered in Sec.~\ref{sec:diss}.

While it is straightforward to see that, in general, $\kappa(\mathbf{M}^{(1)})$ can take any value from 1 (best condition) to infinity (ill-condition), we note that under the best achievable minimum condition number $\kappa(\mathbf{M}^{(1)})$ in linear quantum state tomography is often higher than 1 and is related to the Hilbert dimension $D$. 
It is known that the so-called symmetric informationally complete positive operator-valued measures (SIC-POVMs)~\cite{rene04JournalofMathematicalPhysics} form a set of optimal-frame measurements in linear quantum state tomography~\cite{scot06J.Phys.A:Math.Gen.}. 
Therefore, we can use the condition number associated with SIC-POVMs, which is inherently the minimum achievable condition number, for comparison purposes. More specifically, for the instrument matrix $\mathbf{M}_\mathrm{SIC}^{(1)}$ of the SIC-POVMs, there is $\kappa(\mathbf{M}_\mathrm{SIC}^{(1)})= \sqrt{{D+1}}$, with $D$ being the Hilbert dimension for single-photon HG states. We show the derivation of $\kappa(\mathbf{M}_\mathrm{SIC}^{(1)})$ in \revrep{Appendix~\ref{app:Der}}{ the Appendix}.

\section{Numerical Simulation of Freeform Metasurfaces for Random Tests}\label{sec:sim}

\begin{figure}[t]
\includegraphics[width=0.5\textwidth]{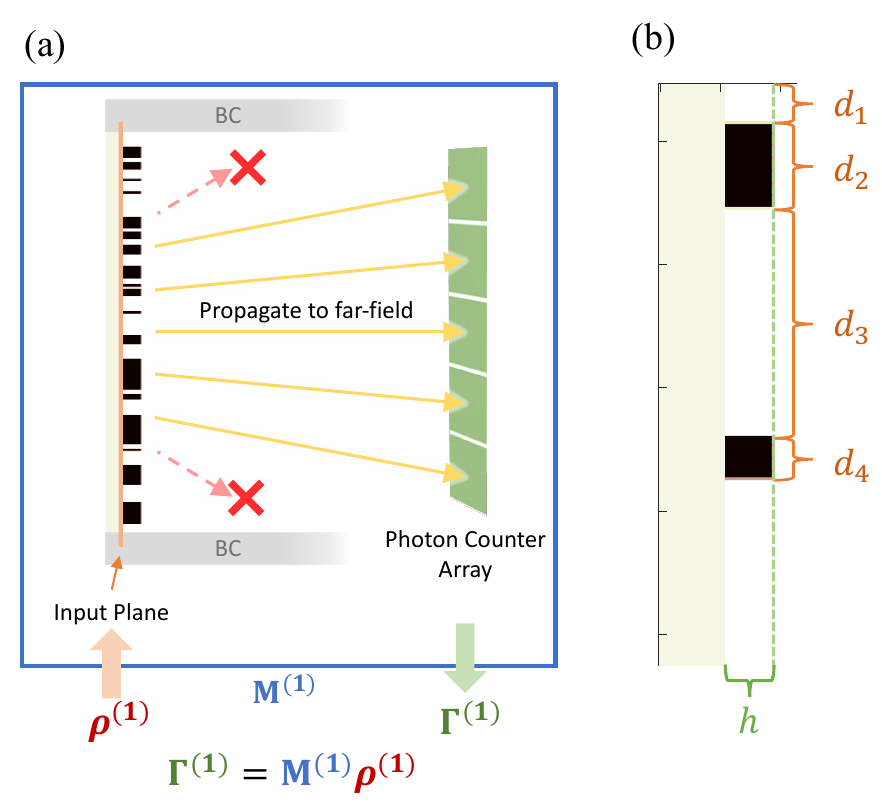}
\caption{\label{fig:design}Numerical experiment setup. (a) The single-photon simulation setup of a metasurface. The single photon input state, $\bm{\rho}^{(1)}$, is scattered by the metasurface and propagated to the far-field detector pixels, resulting the expected photon counts of each pixel. This relation between the flattened input density matrix, $\bm{\rho}^{(1)}$, and the expected average photon count vector, $\bm{\Gamma}^{(1)}$, is linearly related by the instrument matrix, $\mathbf{M}^{(1)}$.
(b) Parametrization of a freeform metasurface structure. A 1-dimensional freeform metasurface can be parametrized by the structure thickness, $h$, and an array, $\mathbf{d}$, representing the widths of the ridges and gaps.}
\end{figure}

In this section, we will describe how we parameterize and simulate the freeform metasurface and how to obtain the single-photon instrument matrix $\mathbf{M}^{(1)}$ from the scattering matrix of the metasurface. 

In Fig.~\ref{fig:design}, we sketch the simulation setting of our metasurface. An overview of the numerical simulation setup is shown in Fig.~\ref{fig:design}(a). The unknown monochromatic input state is illuminated onto the metasurface, and the diffracted output is numerically simulated to propagate through free space, reaching the single-photon detector array at the far field. We assume that such a detector array can count single photons in their probabilities of arriving at each pixel as well as the probability of $N$ photons simultaneously arriving (i.e., $N$-fold correlations).

Our numerical test is based on a freeform configuration of a metasurface as shown in Fig.~\ref{fig:design}(b). For a fixed structure height ($h$), the freeform metasurface is characterized by a parameter vector $\mathbf{d}=[d_1, d_2, d_3, ...]$ representing the widths of the nanoridges (also known as meta-atoms) and the gaps between them. 
Note that, in this freeform parametrization, the number of meta-atoms [i.e., $\revadd{n_\mathrm{atom}=}\left( \mathrm{size}(\mathbf{d})-1\right)/2$] is not always a fixed parameter (see Sec.~\ref{sec:perf} for more details), and the in-plane dimension of the metasurface (i.e., $\sum\mathbf{d}$) is fixed to be around $247$ wavelengths, going beyond the typically-used small, predefined, unit-cell geometry that has limited functionalities. Therefore, our metasurface is a genuine large-scale structure. 

In the frequency domain solver, the boundary condition of this fixed-size structure is set to be periodic with \revadd{sufficiently} large padding \revadd{of $100\ \mathrm{\mu m}$} on both sides, giving numerous discrete reciprocal space ($k$-space) output channels, among which the evanescent channels are then neglected. After free space propagation to the far field, each \revadd{and every} propagating $k$-space channel \revrep{are}{ can be} mapped to a detector pixel for photon counting \revadd{and we assume no photon loss, which would correspond to a unity fill factor}. We simulate fewer available detector pixels ($L$) by binning of $k$-space channels without affecting the unity fill factor.

For any metasurface structure parametrized by $\mathbf{d}$ and $h$, a large-scale frequency-domain simulation of this entire metasurface can reveal the transformation of the probability amplitude of the input photons, represented by $\mathbf{T}$, the transmission matrix block of the scattering matrix, $\mathbf{S}$. This matrix $\mathbf{T}$ of size $L\times D$ represents the linear transformation of the metasurface from the HG states coefficients to the discrete $k$-channel amplitudes. Specifically, for any single-photon pure state $\ket{\psi^{(1)} }$ under the HG basis,
\begin{equation}\label{eq:wavefunction_relation}
    \ket{\phi} = \mathbf{T}\ \ket{\psi^{(1)} },
\end{equation}
with $\ket{\phi}$ being the probability amplitudes at the far-field detector pixels. Hence, the average photon count at the $l$-th pixel can be calculated through the pixel's probability of detecting a photon, which is
$\mathbf{\Gamma}^{(1)}_l=|\braket{l|\phi}|^2,$
where $\bra{l}$ denotes the row vector with all zero elements except for the $l$-th element being 1. Inserting Eq.~\eqref{eq:wavefunction_relation} we have
$
\mathbf{\Gamma}^{(1)}_l=\braket{l|\mathbf{T}|\psi^{(1)}}\braket{\psi^{(1)}|\mathbf{T}^\dagger|l}$
for pure states. Hence, it is straightforward to see that for a density matrix $\hat{\rho}^{(1)}$, we have
\begin{equation}\label{eq:Gamma_l}
\mathbf{\Gamma}^{(1)}_l=\bra{l}\mathbf{T} \hat{\rho}^{(1)} \mathbf{T}^\dagger\ket{l}.
\end{equation}
To obtain the instrument matrix $\mathbf{M}^{(1)}$ that fulfills $\bm{\Gamma}^{(1)} = \mathbf{M}^{(1)} \bm{\rho}^{(1)}$, we can first use Eq.~\eqref{eq:Gamma_l} to identify that $\mathbf{\Gamma}^{(1)}_l=\sum_{u,v} T_{lu}  T_{lv}^\ast \rho^{(1)}_{uv}$ where $u,v$ are matrix indices. Hence, we find
\begin{equation}
    M^{(1)}_{l\ (u,v)}=T_{lu}  T_{lv}^\ast,
\end{equation}
where the $(u,v)$ indexing in $M^{(1)}$ can be accordingly converted to a single index following how $\hat{\rho}^{(1)}$ is flattened into $\bm{\rho}^{(1)}$.
Note that this instrument matrix is only a function of metasurface structure, $\mathbf{M}^{(1)}(\mathbf{d}, h)$, and is independent of the input state. Later, the multiphoton instrument matrix, $\mathbf{M}^{(N)}$, can be simply obtained by applying Eq.~\eqref{tensor_prod_M}.

An enabling factor in our many random tests is the capability to run the large-scale simulations of the entire metasurface. In this work, we employ the MESTI (Maxwell's Equations Solver with Thousands of Inputs) implementation of the Augmented Partial Factorization (APF) method~\cite{lin22NatComputSci} to perform the simulation. The APF is a frequency-domain approach that is particularly suitable for solving the scattering matrix $\mathbf{S}$ associated with far-field responses, as its partial factorization step avoids the direct computing of large matrix inversion by inherently ignoring near-field information. 
Each run of the APF simulation that computes $\mathbf{S}$ of our metasurface structure only takes less than a minute using a desktop computer.  
We hence use the computed $\mathbf{S}$ to obtain its transmission matrix block $\mathbf{T}$, and finally we use the formalisms described above to eventually convert $\mathbf{T}$ to the instrument matrix $\mathbf{M}^{(1)}$.

\section{\label{sec:perf} Performance of Randomly Structured Metasurfaces}

In our work, we choose the metasurface material to be commonly used amorphous silicon ($n=3.74$) on quartz  ($n=1.45$) at a wavelength of $810\ \mathrm{nm}$, which is a typical wavelength for nonlinear-generated photons. The total width $W$ of the freeform metasurface is set to $200\  \mathrm{\mu m}$, being much larger than the zero-th order radius $w_0= 20\ \mathrm{\mu m}$, such that almost all photons will incident the nanostructured region even for high order HG states of interest. The periodic boundary condition on the structure with padding gives a total of $987$ propagating output ports for a maximum of $987$ detector pixels. To explore the performances of random structures, we first identify a few key structural parameters to vary, including ridge height ($h$), average structure size [$\mu=\mathrm{mean}(\mathbf{d})$], number of HG orders or Hilbert dimension ($D$), while keeping other parameters fixed; also, we consider the effect of a minimum fabrication feature size ($d_\text{min}$) constraint of at least $100\ \mathrm{nm}$ feature size, and compare it with the ideal case without any constraint ($d_\text{min}=0$). 

Now, we define what we mean by ``random'' in this paper. For each metasurface we use in our random test, we randomly generate the elements in $\mathbf{d}$ which describe the widths of the \revadd{$n_\mathrm{atom}$} ridges and \revadd{the $(n_\mathrm{atom}+1)$} gaps \revadd{following a specific probability distribution with some tunable parameters. The structure generation is subjected to additional constraints for fabrication considerations that the total structural width is fixed to $W=\sum_i^{2n_\mathrm{atom}+1} d_i = 200\ \mathrm{\mu m}$, and all elements (except for the first and the last gaps) should be larger than a minimum feature size, $d_i\ge d_{\min}$. We focus on studying metasurface $\mathbf{d}$ generated by placing $2n_\text{atom}$ cuts within the free-allocable space after removing the minimum feature sizes $W_\text{alloc}$, and subsequently adding $d_\mathrm{min}$ to each element. The resulting structure $\mathbf{d}$ follows a Dirichlet distribution with each element following a marginal Beta distribution, $d_i \sim W_\mathrm{alloc} \cdot \text{Beta}(1, 2 n_\mathrm{atom}) + d_{\min},$}
\revrep{where $W_\mathrm{alloc} = W - (2n_\mathrm{atom}-1) \cdot d_{\min}$}{ The elements of $\mathbf{d}$ follow a Gaussian distribution. More specifically, it is generated by generating a random number  $r_j \sim \mathcal{N}\left(\mu-d_{\text{min}}, (\mu-d_{\text{min}})^2 \right)$ for the $j$-th element of $\mathbf{d}$ given by $d_j=d_\mathrm{min}+r_i$. Here $d_\text{min}$ is the minimum feature size set by the fabrication constraint, and we term $\mu$ as the average meta-atom size, as there is $\mu= \mathrm{mean}(\mathbf{d})$ for the random parameters generated this way}. \revadd{In the limit of large $n_\mathrm{atom}$ (which is the case for the structures in this paper), this distribution approximates a shifted exponential distribution $\mu\approx\mathrm{std}(\mathbf{d})\approx W/(2n_\mathrm{atom})$, where $\mu\equiv \mathrm{mean}(\mathbf{d})$ is the average meta-atom size and $\mathrm{std}(\cdot)$ is the standard deviation function.}
\revadd{In the result below, we will show $\mu$ as the independent parameter for readability instead of $n_\mathrm{atom}$ which is the actual parameter we tune.}


Then, we run a series of numerical simulations with different heights $h$, Hilbert dimension $D$, and number of \revrep{meta-atoms $n_\mathrm{atom}$ (or equivalently $\mu$)}{ detector pixels $L$}. We also vary the number of detector pixels \revadd{$L$} by binning the propagating $k$ channels. 
For each parameter setting, we calculate the condition numbers of $30$ randomly generated metasurfaces, from which we obtain the averages and standard deviations to evaluate this set of parameters. In total, we simulated and analyzed a total of \revrep{$16,422$}{ 11508} distinct metasurfaces. 

\revadd{In addition, we also checked generation process that create independent and identically distributed (IID) random structures based on Gaussian distribution (see Appendix~\ref{app:Ran} for more details and a comparison with a discrete Poisson-random process). A Gaussian process $\mathcal{N}(\mu_\mathrm{param}, \sigma_\mathrm{param}^2)$ is defined by parameters $\mu_\mathrm{param}$ and $\sigma_\mathrm{param}$ that represent the expected mean and standard deviation of this variable. We define $\sigma_\mathrm{param}$ to be the spatial randomness of a structure as it describes the spread of the feature sizes with $\sigma_\mathrm{param}=0$ being a periodic grating without randomness and $\sigma_\mathrm{param}\rightarrow\infty$ being the uniform distribution. To respect the fabrication constraints, the probability distribution is truncated and renormalized, and structures extend beyond the total width are discarded. Consequently, the number of meta-atoms $n_\mathrm{atom}$ is no longer a fixed parameter in this scheme, and $\mu_\mathrm{param}$ and $\sigma_\mathrm{param}$ can deviates from the actual statistical average $\mu$ and standard deviation $\sigma$ of $\mathbf{d}$. With a selected set of $h$, $D$ and $L$, we vary $\mu_\mathrm{param}$ and $\sigma_\mathrm{param}$. Similar to above, for each set of parameters, we use the statistical average and standard deviation of the condition numbers of $30$ randomly generated metasurfaces as the quality factor of this set of parameters.}

Below, we base our analysis on representative cases among these results.

\begin{figure}[t]
\includegraphics[width=\textwidth]{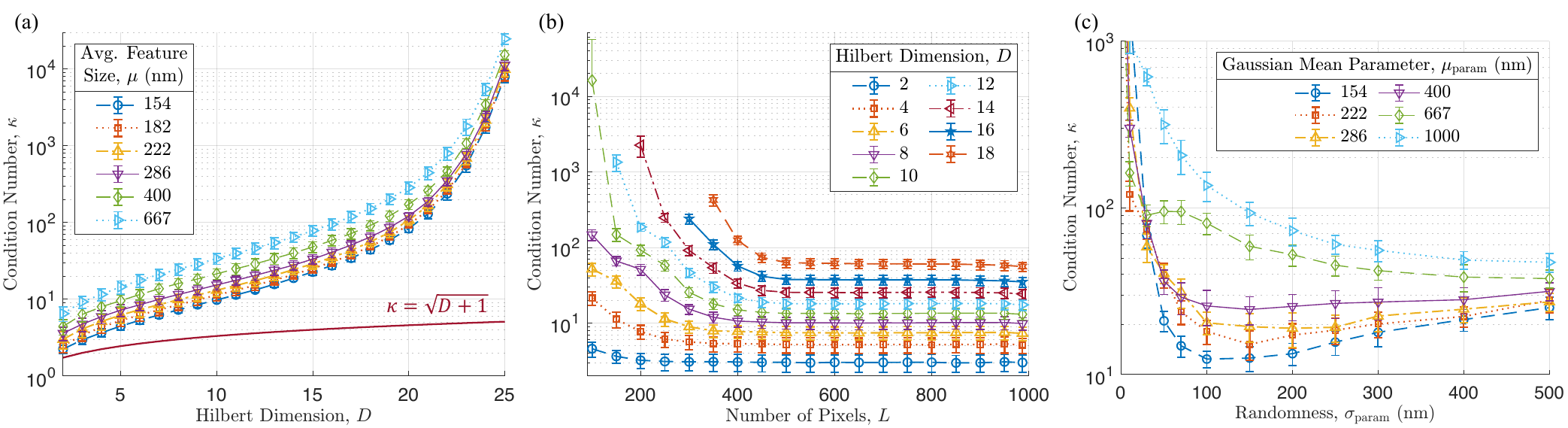}
\caption{\label{fig:random}Effects on condition number, $\kappa$ from (a) different orders of HG states of reconstruction, or equivalently, Hilbert dimension, $D$, \revdel{and} (b) different number of measurement pixels, $L$\revadd{, and (c) structure randomness, $\sigma_\mathrm{param}$}. Here, no minimum feature size constraint is imposed, and the thickness is fixed at $800\ \mathrm{nm}$. In (a), all 987 detector pixels are used; the theoretical condition number of $\kappa(\mathbf{M}_\mathrm{SIC}^{(1)})=\sqrt{D+1}$ is plotted as a red line for comparison. In (b), the average meta-atom size is set to 222 $\mathrm{n m}$; ill-posed points ($L<D^2$) are omitted. \revadd{In (c), all 987 detector pixels are used with Hilbert dimension $D=10$.}
The condition number reduces monotonically with fewer HG states and more measurements\revadd{, and it decreases with structure randomness in general}.}
\end{figure}

\paragraph*{Hilbert Dimension $D$ and the Needed Detector Pixels $L$.} Fig.~\ref{fig:random} \revadd{(a) and (b)} show\revdel{s} the numerical testing results that aim to understand how the detector pixels $L$ and the Hilbert dimension of the HG states $D$ affect the metasurface's performance. 
Specifically, in Fig.~\ref{fig:random}(a), we show the results of the condition number $\kappa$ in logarithmic scale versus different $D$ in the case of a fixed number of detectors $L=987$. 
To show a wide range of metasurface structures, the shown results include several different average meta-atom sizes $\mu$ with a fixed thickness $h=800\ \mathrm{nm}$. Here, there is no constraint on the minimum feature size; i.e., we set $d_\text{min}=0$.
There is a clear fast-growing trend of $\kappa$ in all cases of $\mu$ as $D$ increases, see Fig.~\ref{fig:random}(a); starting from $D\approx 15$, the growth becomes very fast, beyond exponential. As an example, in this configuration, if we aim to achieve $\kappa \leq 10$, with such a number of detectors that use all the propagating $k$ channels, $D$ should be limited to $D\leq 10$.
In Fig.~\ref{fig:random}(b), we show the results with a fixed $\mu=222\ \mathrm{nm}$ [a relatively good case in Fig.~\ref{fig:random}(a)], while the different curves denote different values of $D$ and the horizontal axis is $L$. We see that 
when pixel number ($L$) is reduced to close to $D^2$ (just satisfying the well-posedness), the condition number is huge; see e.g. the data point of $D=10,\ L=100$ where $\kappa \sim 10^4$. 
With increasing $L$, the condition number $\kappa$ decreases fast initially and then barely decreases after the number of pixels becomes sufficiently large compared to the number of unknowns. Such an observation implies that a sufficiently redundant amount of detector pixels compared to the free unknowns in the density matrix is required for a well-conditioned tomography. For example, with $10$ HG orders ($D=10$), there are $D^2=100$ real-valued unknown variables in the density matrix; using $L=200$ detector pixels would yield a condition number of $\kappa \approx 100$, yet with $L=400$ pixels it drops to below $\kappa \approx 15$, becoming practical for state reconstruction.

\paragraph*{\revadd{Randomness of the Structure $\sigma_\mathrm{param}$.}}
\revadd{In Fig.~\ref{fig:random}(c), we use the standard deviation parameter $\sigma_\mathrm{param}$ of the untruncated Gaussian distribution as structural randomness, and we show how condition number changes with it. Fixing $h=800\ \mathrm{nm}$, $D=10$, $L=987$, the structures in the plot are generated from the Gaussian distribution with parameters $\mu_\mathrm{param}$ and $\sigma_\mathrm{param}$ truncated and renormalized between $d_{\min} = 0$ and $d_{\max} = 10~\mathrm{\mu m}$. 
The condition number $\kappa$ explodes at small $\sigma_\mathrm{param}$; and for non-random structures (periodic gratings) with $\sigma_\mathrm{param}=0$, their condition numbers are above $10^6$ from our numerics. On the other hand, $\kappa$ decreases with the randomness $\sigma_\mathrm{param}$ in general. This confirms that the spatial randomness of the metasurface structure indeed creates a tomographically more complete set of measurement operators and thus reducing the condition number.
For some curves, the condition number showing an increasing trend with the randomness for $\sigma_\mathrm{param}>100\ \mathrm{nm}$ is because their actual average meta-atom sizes $\mu$ are increasing with $\sigma_\mathrm{param}$ in the meantime (see Appendix~\ref{app:Ran} for more details).
}

\begin{figure*}
\includegraphics[width=\textwidth]{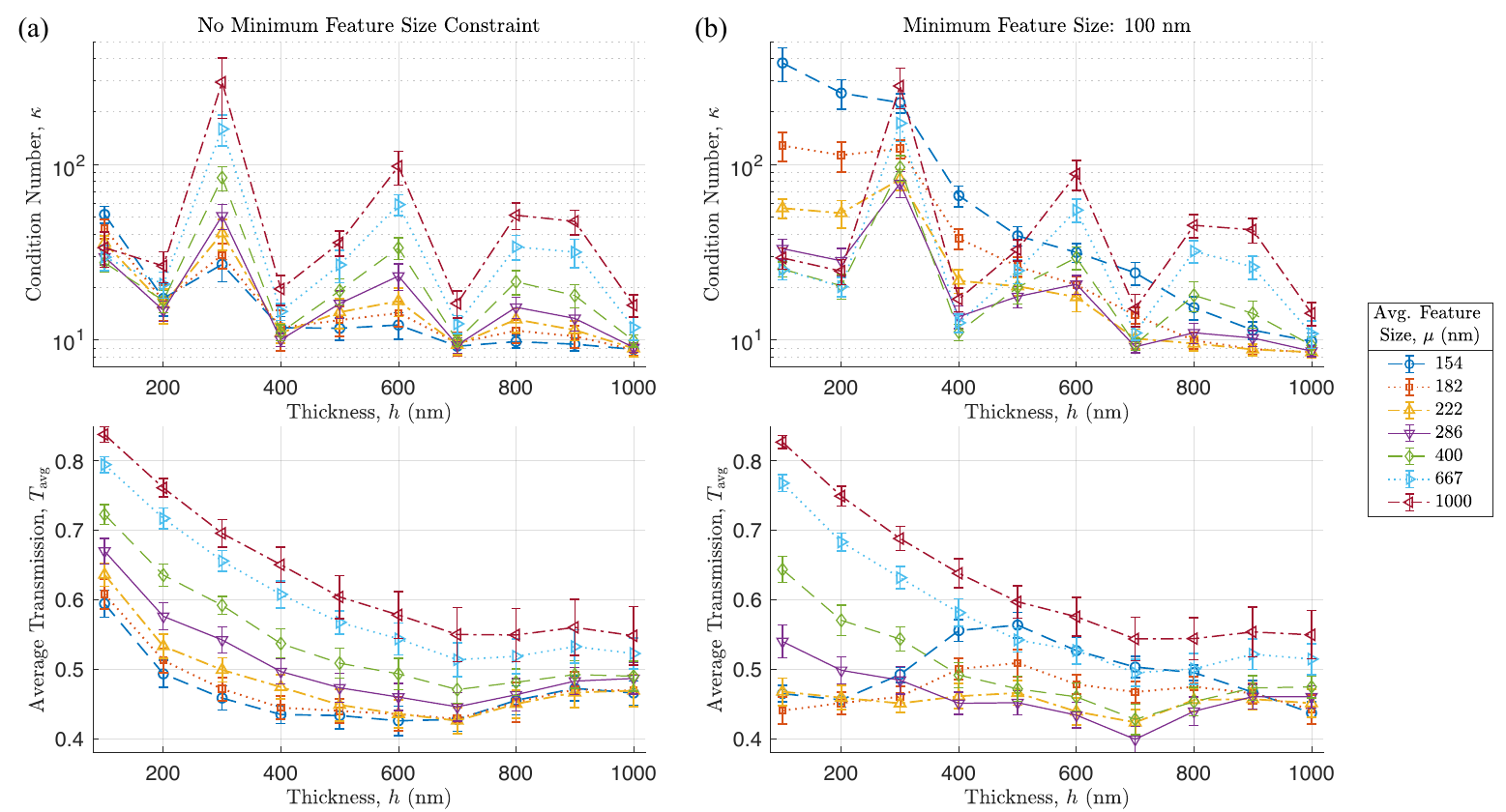}
\caption{\label{fig:random2}Plots of condition number, $\kappa$, and average transmission, $T_{\mathrm{avg}}$, with respect to metasurface thicknesses, $h$. The Hilbert dimension is set to 10 and all 987 detector pixels. The left plots (a) are without minimum feature size constraint $d_\text{min}=0$, and the right plots (b) impose a minimum feature size constraint of $d_\text{min}=100\ \mathrm{n m}$.}
\end{figure*}

\paragraph*{Meta-atom Layer Thickness $h$.}
In Fig.~\ref{fig:random2},  we show the effects of meta-atom thickness $h$ and the minimum structure size $d_\text{min}$ on the device's performance. With increasing $h$, the $\kappa$ exhibits an overall decreasing trend but with fluctuations approximately with a period of the wavelength of light in the meta-atom material. Without minimum feature size constraint [Fig.~\ref{fig:random2}(a, top)], most of the condition numbers tend to be smaller with smaller average atom size $\mu$ for the same $h$.  On the other hand, in Fig.~\ref{fig:random2}(b, top) where a minimum feature size of $d_\text{min}=100\ \mathrm{nm}$ is imposed, there is a dramatic increase of condition numbers when the $\mu$ gets close to $d_\text{min}$ as there is less variety across different meta-atoms.
In addition to $\kappa$, we also plot the average transmission $T_\mathrm{avg}$. We calculated $T_\mathrm{avg}$ by sending in a fully mixed state, which is equivalent to averaging the transmission of all possible input states. We examine $T_\mathrm{avg}$ as maximizing efficiency, which is another important factor of consideration. As can be seen from Figs.~\ref{fig:random2}(a, bottom) and (b, bottom), $T_\mathrm{avg}$ generally decreases with larger $h$ and smaller $\mu$ when $\mu$ is far from $d_\text{min}$.

\paragraph*{Average Meta-Atom Size $\mu$ and Minimum Feature Size $d_\text{min}$.}
The condition number $\kappa$ generally decreases with a smaller average meta-atom ($\mu$) size [see Fig.~\ref{fig:random}(a)] in absence of a minimum feature size constraint, reaching a limit above the theoretical $\kappa(\mathbf{M}_\mathrm{SIC}^{(1)})$ value of $\sqrt{D+1}$. However, with fabrication consideration, when a minimum feature size, $d_\text{min}>0$, is in place, the structure will have less randomness with reducing average size, $\mu$, and gradually converge toward a periodic grating with a periodicity of  $2d_\text{min}$. 
As a result, for each thickness $h$, there exist a minimizer $\mu_\mathrm{opt}$ that minimizes $\kappa$, and this $\mu_\mathrm{opt}$ changes with $h$.

Overall, we find that random metasurface structures are capable to be well-conditioned with appropriate choices of parameters. The number of detectors must be sufficiently redundant, that is, much more than the free real variables in the density matrix. The thickness of structure ideally needs to correspond to a local minimum of the condition number while having a high average transmission; and in general choosing a smaller thickness helps as the transmission is at least maximized. In practice, the structure's thickness, fabrication feature size constraint and detector pixel numbers may not be easily adaptable; so, one key factor is to find the best average meta-atom size, which should be small in general, yet not comparable to the minimum fabrication feature size to enable sufficient room for randomness. 

\section{\label{sec:Rec} Qudit State Reconstruction}

\begin{figure}[t]
\includegraphics[width=0.9\textwidth]{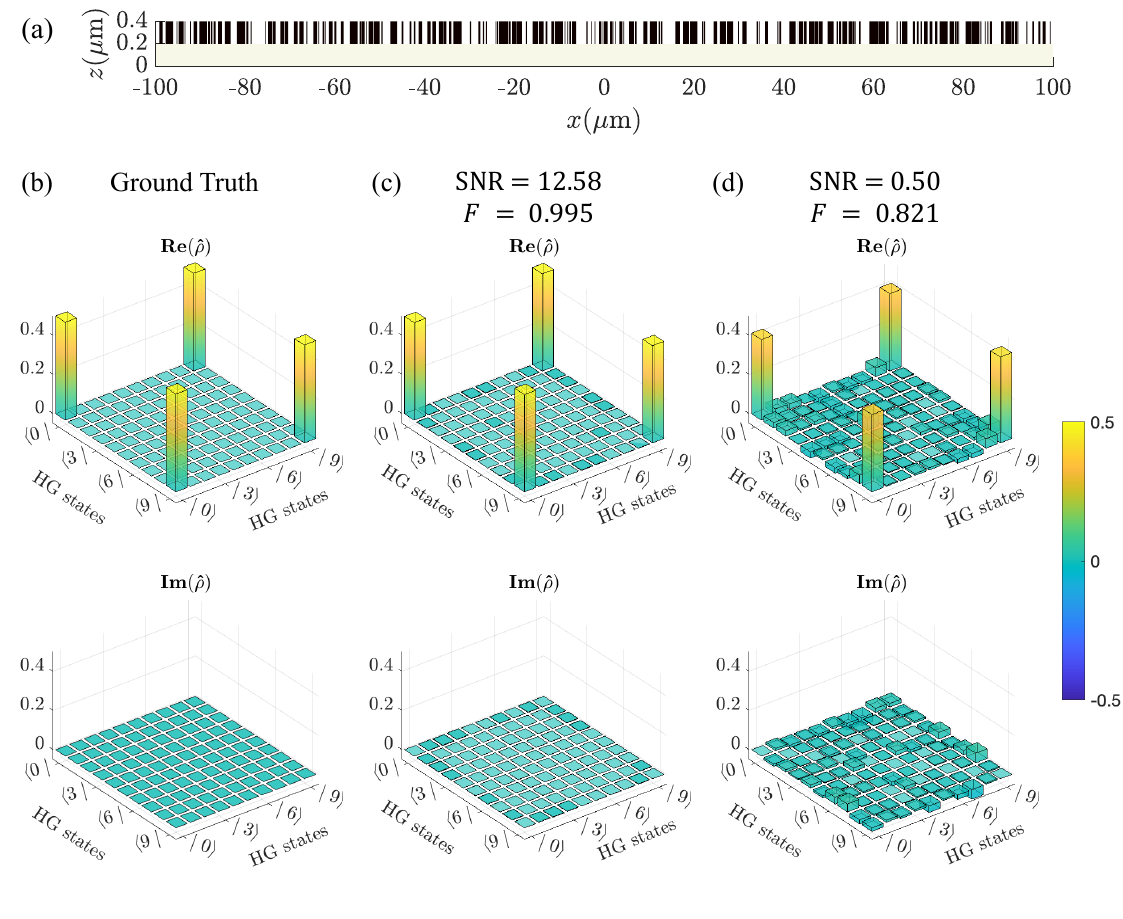}
\caption{\label{fig:single_rec}Reconstruction of an example 10-HG-order input state using a selected random metasurface $h=200\ \mathrm{nm}$, $\mu=667\ \mathrm{nm}$, and $d_\text{min}=100 \ \mathrm{nm}$, whose structure is shown in (a). The ground truth density matrix of the input is shown in (b). With SNRs of 12.58 and 0.50, the reconstructed density matrices are reconstructed to have fidelities of 0.995 as in (c) and 0.821 as in (d) respectively.}
\end{figure}

In this section, we discuss qudit state reconstruction with examples of a specific randomized metasurface selected from the numerical tests. 
We selected a best-performing set of key parameters, considering both the condition number and the average transmission. Specifically, we target the reconstruction of $10$-HG-order states with all $987$ pixels and a fabrication constraint of $d_\text{min}=100 \ \mathrm{nm}$ [see Fig.~\ref{fig:random2}(b)]. Hence we chose the $h=200\ \mathrm{nm}$ thick and $\mu=667\ \mathrm{nm}$ average meta-atom size as our key parameters for the random shape generation for $10$-HG-states reconstruction. We ran $30$ random generations under this setting, and picked one structure with the minimum condition number $\kappa=14.8$ and an average transmission $T_\text{avg}=0.678$. Such a metasurface structure is shown in Fig.~\ref{fig:single_rec}(a). Although the structure is a selected one, we emphasize that the variance of the condition number at these key parameters is minimal and there should be little difference from using a completely random structure at these key parameters. 

With this well-conditioned metasurface, we employ the Maximum Likelihood Estimation (MLE)~\cite{hrad97Phys.Rev.A} to perform qudit state reconstruction. Compared to directly computing pseudoinversion of the instrument matrix for the reconstruction, MLE ensures the reconstructed density matrix is physical, as we can preset the physical conditions, which can include Hermiticity, positive semidefiniteness, probability conservation, and multiphoton permutation symmetry. 
In realistic applications of such metasurfaces, the fabricated metasurface must first be characterized to obtain the instrument matrix, $\mathbf{M}^{(N)}$. Such a characterization can be done by preparing a set of well-chosen, known probe states (classical or single-photon) and recording the detectors' responses. Various experimental imperfections will be automatically accounted for in this process. 

Then, with the measured average photon counts  (for $N=1$) or $N$-fold correlation (for $N>1$), denoted by $\tilde{\bm{\Gamma}}^{(N)}$, the MLE algorithm can be formulated as finding the most likely physical vectorized density matrix input, $\bm{\rho}^{(N)}$, to generate noisy measurement. A physical density matrix would be Hermitian, positive semidefinite and unit-trace; and for indistinguishable multiphoton states, an additional permutation symmetry is required to be obeyed.
Here, we apply a simple likelihood function of $2$-norm, and this can be written as:
\begin{equation}
    \mathrm{argmin}_{\bm{\rho}^{(N)}} \ {\|\tilde{\bm{\Gamma}}^{(N)}-\mathbf{M}^{(N)}\bm{\rho}^{(N)}\|}_2,\  \text{s.t. }  \bm{\rho}^{(N)} \text{ is physical.} \label{MLE}
\end{equation}
As an example, we numerically test the randomly generated structure [Fig.~\ref{fig:single_rec}(a)]. Using this structure, we perform both single- and entangled two-photon reconstruction with noisy input to validate its effectiveness. As a measure of the reconstruction performance, we use the fidelity $F\equiv \left(\operatorname{Tr} \sqrt{\sqrt{\hat{\rho}} \hat{\rho}' \sqrt{\hat{\rho}}}\right)^2 $ between the reconstructed density matrix $\hat{\rho}'$ and the ground truth density matrix $\hat{\rho}$, which ranges between zero and unity, with higher $F$ meaning the two states being more similar to each other~\cite{jozs94J.Mod.Opt.a}.

For single-photon qudit reconstruction, choose an example $D=10$ spatial state input made of equal fraction of the lowest and the highest HG order, $\ket{\psi^{(1)}_1 }=\frac{1}{\sqrt{2}}(\ket{0}+\ket{9})$, for testing; the density matrix of this state is plotted in Fig~\ref{fig:single_rec}(b). 
Using all $L=987$ available pixels in our simulation setup, we obtain the theoretical average photon output at the detector location, $\bm{\Gamma}^{(1)}$, by applying Eq.~\eqref{eq_instrument_matrix_N} after simulating the device's instrument matrix, $\mathbf{M}^{(1)}$. 
Then, we add a random Gaussian noise, $\bm{\epsilon}$, to simulate the various possible sources of noise during measurement, giving the noisy measurement, $\tilde{\bm{\Gamma}}^{(1)}$. The added noise level can be characterized by the SNR as defined in Eq.~\eqref{SNR_meas}.
Lastly, we use this noisy measurement to run the MLE algorithm described in Eq.~\eqref{MLE} for reconstruction. With the reconstructed density matrix, we calculate its fidelity $F$ compared to the ground truth. It is worth noting that this method works for arbitrary pure or mixed states in general, although the example here uses a pure state. 

We first simulate a noisy measurement signal having an SNR of $12.58$ and perform the numerical reconstruction using MLE, and we achieve a nearly unity reconstruction fidelity of $F=0.995$ [Fig~\ref{fig:single_rec}(c)]. Such a high fidelity shows that the metasurface is excellent at reconstructing with moderate noise level. Then, we further test it with a very poor SNR of $0.50$, and the resulting reconstructed density matrix still exhibits a reasonably good fidelity of $F=0.821$ [Fig~\ref{fig:single_rec}(d)], demonstrating robustness of such a randomly shaped metasurface in reconstructing noisy data.

\begin{figure}[t]
\includegraphics[width=0.55\textwidth]{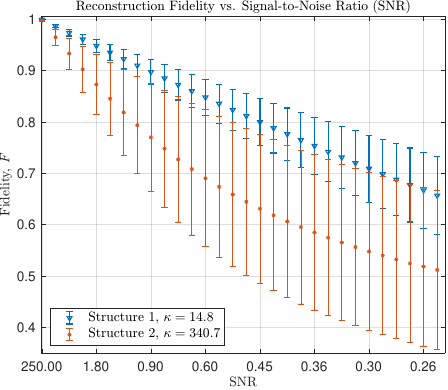}
\caption{\label{fig:fid_vs_noise}Statistical means and standard deviations (error bars) of the reconstruction fidelities vs. decreasing signal-to-noise ratio (SNR). Note the abscissa is not on a linear scale in SNR values, but is on a linear scale with the standard deviation of added noise. Two randomly structured metasurfaces with different condition number, $\kappa$, are tested with $50$ sets of random noise at each SNR, and the average reconstruction fidelities are plotted as data points with the fidelities' standard deviations as error bars.
}
\end{figure}

While we have discussed theoretically how condition number is closely related to robustness in the reconstruction in the presence of noise, here we show two sets of numerical reconstruction tests on this example input using two metasurfaces with different condition numbers. 
In addition to the metasurface shown in Fig~\ref{fig:single_rec}(a), which we call structure 1 for clarity, for comparison, we selected a second metasurface (structure 2) randomly generated at $\mu = 154\ \mathrm{n m}, h=100\ \mathrm{n m}$ exhibiting a much higher condition number of $\kappa = 340.7$, about 23 times that of structure 1 (with $\kappa = 14.8$). Although structure~2 has a lower average transmission, we compensate for this by adding a smaller noise amplitude, ensuring the measurement SNR is the same between the two metasurfaces to isolate the sole effect of the different condition numbers. 
For each SNR level, we ran reconstructions with 50 sets of random noise arrays using both structures, and the average reconstruction fidelities are plotted versus the SNR in Fig.~\ref{fig:fid_vs_noise} with the standard deviations of fidelities shown as error bars.
We find that the fidelities monotonically decrease with increasing noise levels, as shown in Fig.~\ref{fig:fid_vs_noise}. As expected, structure 2's reconstruction fidelity drops quicker than structure 1 as the SNR decreases.

\begin{figure}[t]
\includegraphics[width=0.95\textwidth]{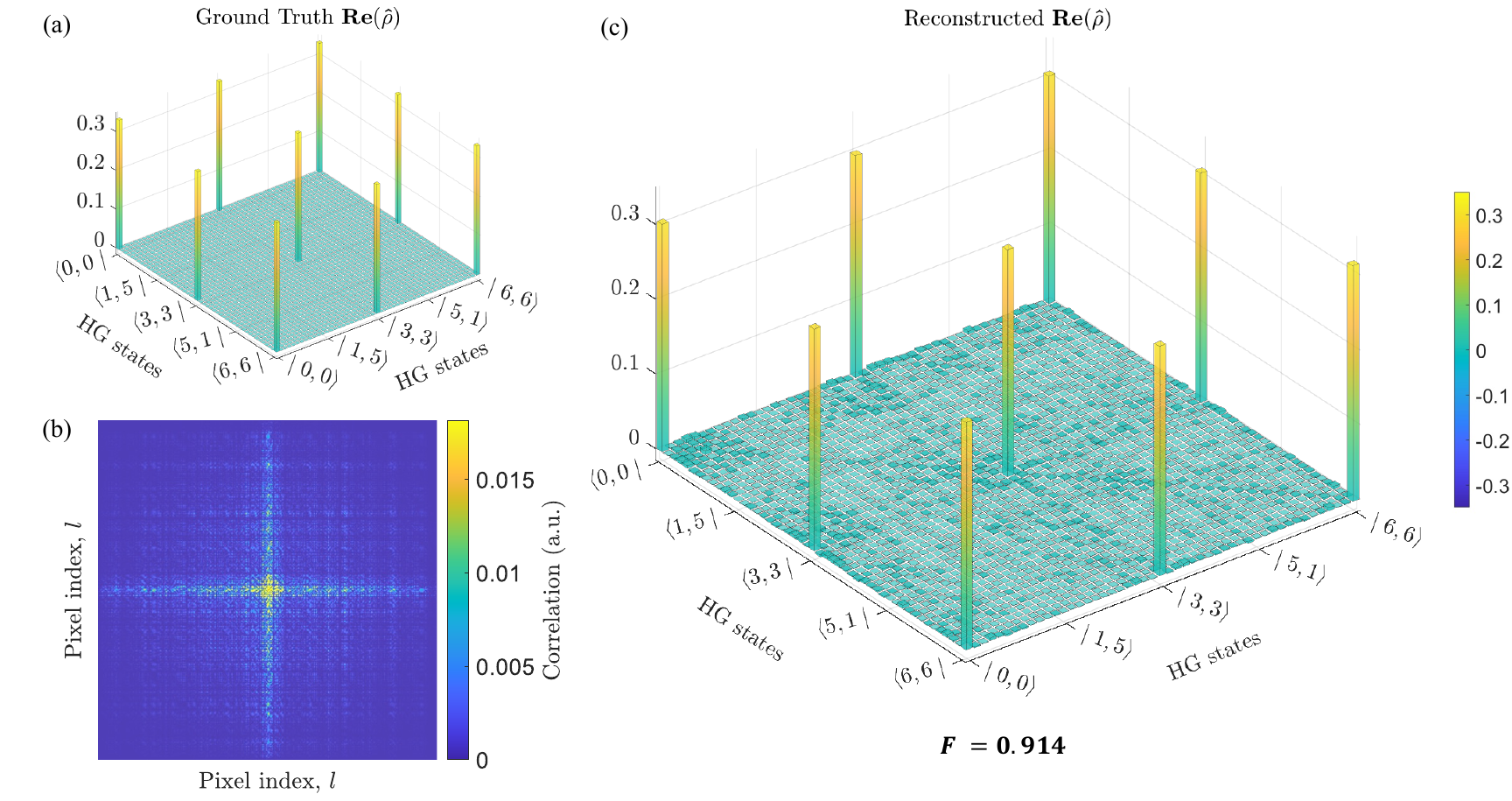}
\caption{\label{fig:bi_rec}Entangled two-photon state reconstruction with a randomly structured metasurface. (a) Density matrix's real part of the example input (imaginary part is zero), $\frac{1}{\sqrt{3}}\left(\ket{0,0}+\ket{3,3}+\ket{6,6}\right)$, a nonseparable two-photon spatial state with $D=7$.
(b) The result noiseless 2-fold correlation image $\bm{\Gamma}^{(2)}$ at the detector location.
(c) The real part of the reconstructed density matrix (its imaginary part is nearly a zero matrix). The reconstruction assumes the use of a click detector (so the diagonal elements of $\bm{\Gamma}^{(2)}$ are excluded) and indistinguishability of the two photons. A random Gaussian noise is added to $\bm{\Gamma}^{(2)}$ resulting an SNR of 0.64. And the reconstruction fidelity $F=0.914$.
}
\end{figure}

Next, based on structure 1, we also show an example reconstruction of an entangled two-photon state with $D=7$, having a joint density matrix of size $49\times49$. We select a nonseparable two-photon state $\ket{\psi^{(2)}_2 }=\frac{1}{\sqrt{3}}(\ket{0,0}+\ket{3,3}+\ket{6,6})$, whose density matrix's real part is shown in Fig.~\ref{fig:bi_rec}(a) and imaginary part is zero. Here, we make an assumption that the two photons are indistinguishable, and the detector cannot resolve photon number. As a result, the simulated ground truth $2$-fold correlation $\bm{\Gamma}^{(2)}$ at the detector \revadd{(Fig.~\ref{fig:bi_rec}b)} is symmetric; and the diagonal elements of $\bm{\Gamma}^{(2)}$, representing two photons hitting one pixel, are removed (see Sec.~\ref{sec:diss} for more detail on the effect of photon-indistinguishability and detector type).
We then added random Gaussian noise with a relatively high amplitude to simulate high measurement noise. We emphasize that applying a simple Gaussian noise model here directly on the correlation data serves as a general example; in realistic correlation measurements, one can adapt the noise model to specific sensors, which is beyond the scope of this work.

The noisy correlation matrix $\tilde{\bm{\Gamma}}^{(2)}$ has a signal-to-noise ratio of $0.64$. With the MLE algorithm in Eq.~\eqref{MLE}, we reconstructed the two-photon density matrix [see Fig.~\ref{fig:bi_rec}(c) for the real part], which yields a decent fidelity of $0.914$ despite the high measurement noise.

\section{\label{sec:diss} Discussion}

\paragraph*{Condition Number versus Full Singular Value Information.}
In this work, the condition number of the metasurface's instrument matrix served as a global indicator of noise amplification, as it bounds the noise amplification  of through reconstruction. While condition number provides a ``worst case scenario'' measure of the inverse condition, it is noteworthy that the full singular values distribution should provide a deeper insight. It is desirable to maximize all singular values in addition to achieving low condition number. Access to all the singular values and singular vectors of the instrument matrix will also provide a means to optimize the metasurface when there is \emph{a priori} knowledge of the measured quantum state, as one can maximize the transmission of the channels associated with the maximum information on the state. 

\paragraph*{Detector Type.}
For multiphoton state reconstruction, the detector type and photon distinguishability are important considerations during state tomography. In contrast to photon-number-resolving detectors assumed in Sec.~\ref{sec:HGQST}, click detectors (e.g., single-photon avalanche photodiodes) are way more commonly used. These click detectors only give binary information, i.e., no photon detected or photon(s) detected. The consequence is that, for the $N$-fold correlation data matrix $\Gamma^{(N)}$, one can only access the fully nonlocal correlations, i.e. the elements $\Gamma^{(N)}_{l_1, l_2, \dots, l_N}$ with indices $l_1, l_2, \dots, l_N$ being all distinct. For $N=2$ case like the example in Fig.~\ref{fig:bi_rec}, using click detector corresponds to removing the diagonal elements of $\Gamma^{(2)}$. For the instrument matrix $ \mathbf{M}^{(N)}$, one needs to remove all its rows that correspond to partially or fully local correlations, obtaining a reduced instrument matrix $\mathbf{M}^{(N)}_\text{cl}$~\cite{titc18npjQuantumInf}. 

\paragraph*{Distinguishability in Photon Detections.}
In Sec.~\ref{sec:HGQST} we assumed for the detectors the photons are distinguishable by some other degree of freedom for simplicity. If the photons are not distinguishable by the detectors, the density matrix is subject to additional symmetry constraints. Taking two-photon states in $D$-dimensional Hilbert space as an example, we can denote their (distinguishable-case) density matrices $\hat{\rho}_{mn,ij}$ using two pairs of indices, where the first pair of indices, $m,n$, are the HG states of the first photon, and the second pair, $i,j$, are those of the second photon. In addition to the Hermiticity requirement of $\hat{\rho}_{mn,ij}=\hat{\rho}^*_{nm,ji}$, the two-photon permutation symmetry due to non-distinguishable detection would also require $\hat{\rho}_{mn,ij}=\hat{\rho}_{ij,mn}$. Combining the two symmetries, it would result in only $\frac{1}{2}D^2(D^2+1)$ free real variables in the density matrix without considering normalization. On the detector side, since one cannot identify which pixel is triggered by which photon, the correlation image is also symmetric with $\Gamma^{(2)}_{l_1,l_2}=\Gamma^{(2)}_{l_2,l_1}$.
If one would like to get a more precise calculation of condition number in this case, one can also remove the columns and rows in $\mathbf{M}^{(N)}$ that are redundant as a result of this permutation symmetry to obtain $\mathbf{M}^{(N)}_\text{ind}$. It is also possible to combine both considerations to obtain a reduced instrument matrix considering both click detectors and photon non-distinguishable photon detection scheme, which we term it as $\mathbf{M}^{(N)}_\text{cl,ind}$.

\paragraph*{Condition Number of Instrument Matrices after Considering Click Detectors or Photon Indistinguishability.}
Removing certain columns and rows in $\mathbf{M}^{(N)}$ to account for click detector and the indistinguishability of photons would indeed result in a different condition number, i.e., in general for $N>1$ we expect $\kappa (\mathbf{M}^{(N)}_\text{cl})$, $ \kappa(\mathbf{M}^{(N)}_\text{ind})$, $\kappa(\mathbf{M}^{(N)}_\text{cl,ind}) $ and $ \kappa^N(\mathbf{M}^{(1)})$ to be pairwise distinct. Nevertheless, for small $N$ and redundant number of detectors $L$, $\kappa(\mathbf{M}^{(1)})$ should still be a good approximate measure of the inverse condition even if we are using click detectors or indistinguishable photons. 

\paragraph*{Experimental Considerations.}
In selecting the random metasurface parameters for numerical testing, we have deliberately chosen the values that are compatible with current nanofabrication capabilities. The choice of amorphous silicon on quartz at $810\ \mathrm{nm}$ ensures both high refractive index contrast and compatibility with established processes such as plasma-enhanced chemical vapor deposition, followed by electron-beam lithography and reactive-ion etching. The metasurface lateral size of $200\ \mathrm{\mu m}$, ridge heights in the simulated range, and minimum feature sizes above $100\ \mathrm{nm}$ are well within the achievable limits of many nanofabrication laboratories and commercial foundries, while still allowing for sufficient structural randomness to guarantee tomographic completeness. After fabrication, the metasurfaces can be integrated \revdel{directly} in front of a detector array \revadd{through a 2$f$ setup with a relatively high numerical-aperture objective}\revrep{. The detector array could be}{ , such as} a scientific-grade electron multiplying charge-coupled device (EMCCD) or single-photon avalanche diode (SPAD) array, which can provide the required spatial resolution \revadd{high fill factor }and low-noise performance for single-photon counting or multiphoton correlation measurement~\cite{defi18Phys.Rev.Lett.}. The large number of available output ports in our design (up to 987 in the present simulations) remains well matched to the pixel counts of existing single-photon-sensitive image sensors, enabling a highly practical experimental transition from simulation.

\paragraph*{\revadd{Robustness against Fabrication Errors.}} \revadd{The robustness against fabrication errors is built into the random structure generation process. In both Gaussian and Poisson distributions (See Appendix~\ref{app:Ran}), the error bars of the condition number are small compared to the average condition number for the many copies of structures generated, which indicates that if the fabrication process can be modeled as random perturbations to meta-atom and gap widths following common distributions like Gaussian or Poisson, the condition number stays decent for the tomography task.}

\section{Conclusion}

As a conclusion, through large-scale simulation of \revrep{$16,422$}{ 11508} distinct metasurfaces each of over 200 wavelengths in size, we show that randomly generated metasurfaces can achieve reliable quantum state tomography of high-dimensional spatial states when certain conditions are met. We summarize such conditions to three points: first, the number of detector pixels must be sufficiently redundant compared to the number of free real parameters in the density matrix. Second, the meta-atom layer thickness needs to be ideally chosen to correspond to a local minimum in the condition number while maintaining average high transmission, which typically favors thinner structures. Third, the average meta-atom size, being the most experimentally-adjustable parameter, should be small but not comparable to the minimum feature size in order to retain sufficient structural randomness and guarantee tomographic completeness of the generalized measurements. 

We detailed the theory and method for performing single- and multiphoton state tomography with a well-conditioned metasurface, as well as practical considerations such as detector type and photon distinguishability. As an example, we selected a random structure among the ones illustrated in Sec.~\ref{sec:perf}  targeting 10‑dimensional single-photon qudit state reconstruction with all 987 detector pixels. This random structure's single-photon instrument matrix has a low condition number of $14.8$ and a good average transmission of $0.678$. In 10‑dimensional single-photon tomography, this example structure delivers an excellent reconstruction fidelity of $0.995$ under an SNR of $12.58$; and under a poor SNR of only $0.50$, the reconstruction fidelity of $0.821$ showcases the noise-robustness of the structure. With a 7-HG-order nonseparable entangled two-photon state (49-dimensional Hilbert space), we show an excellent reconstruction fidelity of $0.914$ under an SNR of $0.64$ using the same metasurface. The results validate that well-conditioned metasurface can indeed function as an effective qudit state tomography platform.

This work not only establishes the feasibility of using spatially randomized metasurfaces for quantum tomography of spatial qudits, but also presents a practical and resource-efficient alternative to traditional inverse-design approaches. By requiring less computational effort during design and exhibiting intrinsic tolerance to small fabrication errors, this method provides a scalable pathway toward miniaturized, robust, and experimentally accessible quantum measurement systems.

\section{Appendix}
\subsection{\label{app:Der} Derivation of the theoretical condition number for SIC-POVM} 

This appendix derives the condition number of the single-photon instrument matrix $\mathbf{M}_\mathrm{SIC}^{(1)}$ composed of symmetric informationally complete, positive operator-valued measures (SIC-POVMs). For simplicity, we omit the photon number superscripts from here.
Consider an optical system that maps an $D$-dimensional density matrix $\hat{\rho}$ to $L$ intensity measurements with $L\ge D^2$. The vectorized density matrix $\bm{\rho} = \mathrm{vec}(\hat{\rho}) \in \mathbb{C}^{D^2}$ is transformed by the instrument matrix $\mathbf{M}_\mathrm{SIC} \in \mathbb{C}^{L\times D^2}$ to yield non-negative, real measurement outcomes $\bm{\Gamma} \in \mathbb{R}^L_{\ge0}$:
\begin{equation}
\bm{\Gamma} = \mathbf{M}_\mathrm{SIC} \bm{\rho}.
\end{equation}
The condition number $\kappa(\mathbf{M})_\mathrm{SIC}$ follows:
\begin{equation}
\kappa(\mathbf{M}_\mathrm{SIC}) = \frac{\sigma_{\mathrm{max}}(\mathbf{M}_\mathrm{SIC})}{\sigma_{\mathrm{min}}(\mathbf{M}_\mathrm{SIC})},
\end{equation}
where $\sigma_{\mathrm{max}}$ and $\sigma_{\mathrm{min}}$ denote the largest and smallest singular values of $\mathbf{M}_\mathrm{SIC}$.

A Symmetric Informationally Complete, Positive Operator-Valued Measure (SIC-POVM)~\cite{rene04JournalofMathematicalPhysics} in dimension $D$ provides the optimal measurement configuration. 
It consists of $D^2$ rank-1 projection operators $\{\hat{\Pi}_i\}_{i=1}^{D^2}$ satisfying:
\begin{equation}
\mathrm{Tr}(\hat{\Pi}_i \hat{\Pi}_j) = \frac{D\delta_{ij} + 1}{D+1}, \quad \mathrm{Tr}(\hat{\Pi}_i) = 1,
\end{equation}
where $\delta_{ij}$ is the Kronecker delta.
The POVM elements $\hat{F}_i = \frac{1}{D} \hat{\Pi}_i$ fulfill $\sum_{i=1}^{D^2} \hat{F}_i = \mathbf{I}$ with $\mathbf{I}$ being the identity matrix in $D$ dimension. For this SIC-POVM, $\mathbf{M}_\mathrm{SIC}$ has rows corresponding to $\mathrm{vec}(\hat{F}_i)^\dagger$ for $i\in\{1,...,D^2\}$, yielding $L = D^2$ measurements:
\begin{equation}
\Gamma_i = \mathrm{Tr}(\hat{F}_i \hat{\rho}).
\end{equation}

The Gram matrix $\mathbf{G} = \mathbf{M}_\mathrm{SIC}^\dagger \mathbf{M}_\mathrm{SIC} \in \mathbb{C}^{D^2 \times D^2}$ has entries given by:
\begin{equation}
\mathbf{G}_{ij} = \langle \mathrm{vec}( \hat{F}_i), \mathrm{vec}( \hat{F}_j) \rangle = \mathrm{Tr}( \hat{F}_i  \hat{F}_j),
\end{equation}
where we utilize the Hermiticity of $ \hat{F}_i$. Substituting the SIC-POVM properties yields:
\begin{equation}
\mathbf{G}_{ii} = \mathrm{Tr}( \hat{F}_i^2) = \frac{1}{D^2} \quad \text{and} \quad \mathbf{G}_{ij} = \mathrm{Tr}(\hat{F}_i  \hat{F}_j) = \frac{1}{D^2(D+1)} \quad (i \neq j).
\end{equation}
This admits a scaled equicorrelation structure:
\begin{equation}
\mathbf{G} = \frac{1}{D^2} \left[ \mathbf{I} + \frac{1}{D+1} (\mathbf{J} - \mathbf{I}) \right],
\end{equation}
where $\mathbf{I}$ is the identity matrix and $\mathbf{J}$ is the all-ones matrix. The eigenvalues of $\mathbf{G}$ are:
    \begin{align}
    \lambda_1 & = \frac{1}{D} && \text{(multiplicity $1$)}, \qquad
    \\
    \lambda_k & = \frac{1}{D(D+1)} &  \forall \ k\in\{2, ..., D^2\} \quad &\text{(multiplicity $D^2 - 1$)}.
    \end{align}

The singular values of $\mathbf{M}_\mathrm{SIC}$ follow as $\sigma_k = \sqrt{\lambda_k}$:
\begin{equation}
\sigma_{\mathrm{max}} = \sqrt{\lambda_1} = \frac{1}{\sqrt{D}}, \qquad 
\sigma_{\mathrm{min}} = \sqrt{\lambda_k} = \frac{1}{\sqrt{D(D+1)}} \quad (k \geq 2).
\end{equation}
The condition number is therefore:
\begin{equation}
\kappa(\mathbf{M}_\mathrm{SIC}) = \frac{\sigma_{\mathrm{max}}}{\sigma_{\mathrm{min}}} = \sqrt{D+1}.
\label{eq:min_kappa}
\end{equation}

\subsection{\label{app:Ran} \revadd{Random structures from different probability distributions}}

\begin{figure}[ht!]
\centering
\includegraphics[width=0.55\textwidth]{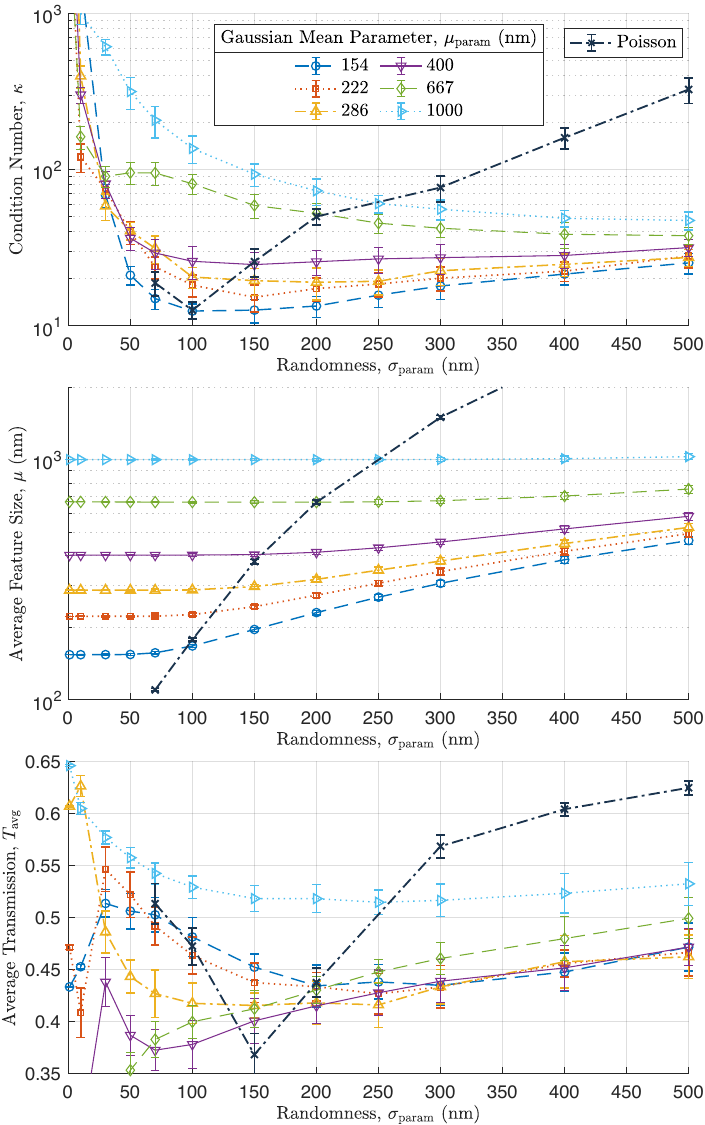}
\caption{\revadd{Condition number $\kappa$ (top), actual average feature size $\mu$ (middle), and the average transmission $T_\mathrm{avg}$ (bottom) as a function of randomness $\sigma_\mathrm{param}$. Gaussian structure curves with different mean parameters $\mu_\mathrm{param}$ are plotted in different colors and the discrete-Poisson structure curve with step size $\Delta x$ and $\mu_\mathrm{param} = \sigma_\mathrm{param}^2/\Delta x$ are plotted as the black dash-dotted curve.}}
\label{fig:poisson}
\end{figure}

\revadd{To study the structures generated from different random processes, we test structures generated from Gaussian probability density distribution $\mathcal{N}(\mu_\mathrm{param}, \sigma_\mathrm{param}^2)$
and discrete Poisson distribution $\mathrm{Pois}(\mu_\mathrm{param})$ with parameters $\mu_\mathrm{param}$ and $\sigma_\mathrm{param}$.}
\revadd{The Gaussian distribution has a probability density function of}
\begin{equation}
\revadd{P_\mathrm{Gaus}(d_i) = \frac{1}{\sigma_\mathrm{param}\sqrt{2\pi}} e^{-\frac{1}{2}\left(\frac{d_i-\mu_\mathrm{param}}{\sigma_\mathrm{param}}\right)^2},}
\end{equation}
\revadd{and the Poisson distribution has a probability mass function of}
\begin{equation}
\revadd{P_\mathrm{Pois}(d_i = k \Delta x) = e^{-(\mu_\mathrm{param}/\Delta x)} \frac{(\mu_\mathrm{param}/\Delta x)^k}{k!}, \quad k = 0, 1, 2, \ldots}
\end{equation}
\revadd{where $\Delta x$ is the discretization step size. The expected average and standard deviation of this Poisson distribution are $\mu_\mathrm{param}$ and $\sigma_\mathrm{param} = \sqrt{\mu_\mathrm{param}\Delta x}$.}

\revadd{In order to respect the fabrication constraint and remove structures that are too large,
we bound the structure size between $d_{\min}$ and $d_{\max}$ by truncating the distributions and re-normalizing them. After truncation, the parameters $\mu_\mathrm{param}$ and $\sigma_\mathrm{param}$ no longer reflects the statistical average and standard deviation, and the number of meta-atoms is no longer fixed due to the total structure width constraint, and the real average size is $\mu = W/(2n_\mathrm{atom}+1)$.}


\revadd{With fixed $h=800\ \mathrm{nm}$, $D=10$, $L=987$, we set the additional structures with constraints of $d_{\min} = 0$ (no minimum feature size constraint) and $d_{\max} = 10~\mu$m. For the Gaussian-random structures, we test the condition number with varying level of randomness $\sigma_\mathrm{param}$ with different Gaussian mean parameters $\mu_\mathrm{param}$; and for the Poisson-random structures, we discretize the structure with a step size of $\Delta x =60\ \mathrm{nm}$ and then we vary $\mu_\mathrm{param} = \sigma_\mathrm{param}^2/\Delta x$. The result is plotted in Fig.~\ref{fig:poisson} \revadd{(top)}, together with plots of the actual average meta-atom size $\mu$ \revadd{(Fig.~\ref{fig:poisson} middle)} and transmission \revadd{$T_\mathrm{avg}$} \revadd{(Fig.~\ref{fig:poisson} bottom)} as a function of the randomness, $\sigma_\mathrm{param}$.}

\revadd{In addition to the summary in Sec.~\ref{sec:perf}, the top two plots in Fig.~\ref{fig:poisson} imply that the condition number of the Gaussian structures increases if the actual average meta-atom size increases. This is consistent with the conclusions on the effect of average meta-atom size $\mu$ in  Sec.~\ref{sec:perf}, and shows that the increasing trend of the condition number $\kappa$ is related to increasing $\mu$. For the Poisson-random structures curve, the condition number is also greatly affected by the increasing $\mu$.}

\revadd{The Poisson random structures at $\sigma_\mathrm{param} = 200~\mathrm{nm}$ have $\mu_\mathrm{param} = 667~\mathrm{nm}$ coinciding with the Gaussian random structures with the same $\mu_\mathrm{param}$ at the same level of randomness in condition number, average meta-atom size, and transmission plots. Comparing with the Beta-distribution-based random structures in Sec.~\ref{sec:perf}, Fig.~\ref{fig:random2} (a), the Beta-distribution-based structures with $\mu=400\ \mathrm{nm}$ have a $\sigma \approx 400\ \mathrm{nm}$, and at $h=800 \ \mathrm{nm}$, who has average condition number of about $20$; this is similar to the condition number of the Gaussian-distribution-based random structures in Fig.~\ref{fig:poisson} (a) with the same level of randomness ($\sigma = 400\ \mathrm{n m}$) and a similar average meta-atom size ($\mu_\mathrm{param} = 400\ \mathrm{nm}$ and $\mu\approx500\ \mathrm{nm}$)
an average condition number of about $29$. This shows that discretization and the random distribution themselves are not major factors affecting the performance of the metasurface.}


\section*{Funding Sources}

This work was funded by the Minist\`ere de l’\'Economie, de l’Innovation et de l’\'Energie (MEIE) of Qu\'ebec, Natural Sciences and Engineering Research Council of Canada (NSERC) [RGPIN-2023-03630, ALLRP 587602-23, ALLRP 580946-22], and Fonds de recherche
du Qu\'ebec - Nature et technologie (FRQ-NT) [\'Etablissement de la rel\`eve professorale, 344951; Regroupement Stratégique Regroupement, Institut transdisciplinaire d'information quantique (INTRIQ)
 and Regroupement Québécois sur les Matériaux de Pointe (RQMP)].
A part of the computations were performed on the digital research infrastructure provided by the Digital Research Alliance of Canada.

\section*{Notes}
This is the author accepted manuscript (AAM) of the article:
[Y. Niu, K. Wang, ACS Photonics (2025), DOI: 10.1021/acsphotonics.5c01909].
The final published version is available at: \url{https://doi.org/10.1021/acsphotonics.5c01909}


\renewcommand{\bibsection}{\section*{References}}
\bibliographystyle{apsrev4-2}
\bibliography{main}








\end{document}